\documentclass[12pt]{iopart}

\usepackage{graphicx}
\begin{document}

\title{Local structure of $RE$FeAsO ($RE$=La, Pr, Nd, Sm) 
oxypnictides studied by Fe K-edge EXAFS}

\author{A. Iadecola$^{1}$, S. Agrestini$^{2}$, M. Filippi$^{3}$, 
L. Simonelli$^{4}$, M. Fratini$^{1}$, B. Joseph$^{1}$, D. Mahajan$^{5,6}$ 
and N.L.Saini$^{1}$} 
\address{$^{1}$ Dipartimento di Fisica, Universit\`{a} di Roma ``La 
Sapienza", P. le Aldo Moro 2, 00185 Roma, Italy}
\address{$^{2}$ Laboratoire CRISMAT, CNRS UMR 6508, ENSICAEN, Boulevard du 
Marechal Juin, 14050 Caen, France}
\address{$^{3}$ Dept of Physics and Astronomy, Vrije Universiteit, 
De Boelelaan 1081, 1081 HV Amsterdam, The Netherlands} 
\address{$^{4}$ European Synchrotron Radiation Facility, 6 RUE Jules 
Horowitz BP 220 38043 Grenoble Cedex 9 France} 
\address{$^{5}$ Department of Materials Science and Engineering, State
University of New York at Stony Brook, Stony Brook, NY 11794, USA} 
\address{$^{6}$ Brookhaven National Laboratory, Upton, NY 11973, USA}

\begin{abstract}
 
Local structure of REOFeAs (RE=La, Pr, Nd, Sm) system has been
studied as a function of chemical pressure varied due to different
rare-earth size.  Fe K-edge extended X-ray absorption fine
structure (EXAFS) measurements in the fluorescence mode has permitted
to compare systematically the inter-atomic distances and their mean
square relative displacements (MSRD).  We find that the Fe-As bond
length and the corresponding MSRD hardly show any change, suggesting
the strongly covalent nature of this bond, while the Fe-Fe and Fe-RE bond
lengths decrease with decreasing rare earth size. The
results provide important information on the atomic correlations that
could have direct implication on the superconductivity and magnetism
of REOFeAs system, with the chemical pressure
being a key ingredient.

\end{abstract} 


\maketitle

The recent discovery of high T$_{c}$ superconductivity in the LaOFeAs
\cite{Kamihara} has triggered intensive research activities on REOFeAs
(RE=rare earth) oxypnictides, producing a large number of publications
focusing on different aspects of these materials
\cite{JPSJIssue,NewJP,Izyumov}. One of the interesting aspects of
these materials is the competing spin density wave (SDW) and the
superconductivity \cite{JZhaoNatP,HuangPRB,JZhaoPRB}.  Indeed, the
undoped compound REOFeAs is antiferromagnetically ordered (albeit a
poor metal), and shows a structural phase transition
\cite{JZhaoNatP,HuangPRB,JZhaoPRB,DCruzNat,Fratini,Margadonna,Karolina}.
 With doping, the system gets superconducting and the structural
transition as well as the SDW transition disappears \cite{JPSJIssue,
NewJP,Izyumov}.  In addition, while the maximum T$_{c}$ of the doped
system increases with reducing the rare-earth ion size
\cite{JPSJIssue,NewJP,Izyumov,REnEPL,CHLee}, the structural
transition temperature decreases for the undoped system
\cite{JZhaoNatP,HuangPRB,JZhaoPRB,DCruzNat,Fratini,Margadonna,Karolina}.
 These observations show interesting interplay between structure,
magnetism and superconductivity with the chemical pressure and
structural topology being important parameters.

It is known that a mere knowledge of the long range ordered structure
is generally insufficient to describe electronic functions of a system
with interplaying electronic degrees of freedom.  Indeed, this has
been shown for transition metal oxides in which the electronic
functions like superconductivity, colossal magneto resistence and
metal insulator transitions are related with interplaying charge- spin
Ð lattice degrees of freedom \cite{StrBond}.  Therefore, a detailed
knowledge of the atomic structure for the REOFeAs oxypnictides could
be a timely feedback to the theoretical models for correlating
structure, magnetism and superconductivity in these materials.

Extended X-ray absorption fine structure (EXAFS) is a site selective
method, providing information on the local atomic distribution around
a selected absorbing atom through photoelectron scattering
\cite{Konings}.  Recently, Zhang et al \cite{Oyanagi} have studied
local structure of doped and undoped LaOFeAs system by Fe K-edge and
As K-edge measurements, providing a temperature dependent anomaly in
the Fe-As correlations at low temperature.  This study was followed by
Tyson et al \cite{Tyson} measuring the same system using Fe K-edge
with no evidence of such anomalies.  Here we have decided to address
a different aspect and exploited the Fe K-edge EXAFS to explore the
local structure of REOFeAs with varying rare-earth size (RE=La (1.16
\AA), Pr(1.13 \AA), Nd(1.11 \AA), Sm(1.08 \AA)).  The results reveal
strongly covalent nature of the Fe-As bond, showing hardly any change
with rare earth size, while the Fe-Fe and Fe-RE bonds show a
systematic size dependence.  On the other hand, the mean square
relative displacements (MSRD) determined by the correlated Debye
Waller (DW) factors of the Fe-Fe bond length decrease with decreasing
rare-earth size and that of Fe-RE seems to increase.  Again, the MSRD
of Fe-As bond remains almost unchanged with the chemical pressure,
underlining the stiffness of this bond.

Fe K-edge X-ray absorption measurements were performed on powder
samples of REOFeAs (RE=La, Pr, Nd, Sm) prepared using solid state
reaction method \cite{RenEPL2}.  Prior to the absorption measurements,
the samples were characterized for the phase purity and the average
structure by X-ray diffraction measurements \cite{Fratini}.  The X-ray
absorption measurements were made at the beamline BM29 of the European
Synchrotron Radiation Facility (ESRF), Grenoble, where the synchrotron
radiation emitted by a bending magnet source at the 6 GeV ESRF storage
ring was monochromatized using a double crystal Si(311) monochromator. 
The Fe K$_{\alpha}$ fluorescence yield was collected using a multi-element
Ge detector array.  Simultaneous transmission signal was measured to
make sure the observed signal to represent true X-ray absorption,
however, it was not possible to obtain absorption signal in
transmission mode without a contribution of the rare-earth
L$_{I}$-edge (6.267KeV, 6.835 KeV, 7.126 KeV and 7.737 KeV
respectively for the La, Pr, Nd and Sm with respect to the Fe K-edge
at 7.112 KeV) and hence the choice was to opt for the partial
absorption signal measured by fluorescence detection for a systematic
comparison.  The samples were mounted in a continuous flow He cryostat
to perform the measurements at low temperature (40 K).  The sample
temperature was controlled and monitored within an accuracy of $\pm$1
K. Several absorption scans were measured to ensure reproducibility of
the spectra and a high signal to noise ratio.  Standard procedure was
used to extract the EXAFS signal from the absorption spectrum
\cite{Konings}, followed by the X-ray fluorescence self-absorption
correction before the analysis.

Figure 1 shows Fe K-edge EXAFS oscillations of REOFeAs samples at low
temperature (40 K) extracted from the X-ray absorption spectra
measured on the powder samples.  The EXAFS oscillations are weighted
by k$^{2}$ to highlight the higher k-region.  There are evident
differences between the EXAFS oscillations due to differing local
structure of REOFeAs with different RE atom (see e.g. the oscillation
around k=6-8 $\AA^{-1}$ and in the k range above $\sim$10-14
$\AA^{-1}$).  The differences in the local structure could be better
appreciated in the Fourier transforms of the EXAFS oscillations
providing real space information.

Figure 2 shows magnitude of the Fourier transforms,
$|$FT(k$^{2}\chi$(k))$|$.  The Fourier transforms are not corrected
for the phase shifts due to the photoelectron back-scattering and
represent raw experimental data.  The main peak at $\sim$2.4 $\AA$ is
due to Fe-As (4 As atoms at a distance $\sim$2.4 $\AA$) and Fe-Fe (4
Fe atoms at a distance $\sim$ 2.8 $\AA$) bond lengths, while the peak
at $\sim$ 3.6 $\AA$ corresponds to the Fe-RE bond length (4 RE atoms
at a distance $\sim$ 3.72 $\AA$).  While the main Fourier transform
peak at $\sim$2.4 $\AA$ appears to shift towards higher R-values, the
Fe-RE peak seems to appear with a decreased amplitude with decreasing
rare-earth size.  The evident shift of the main peak is due to
increased amplitude of the Fe-Fe scattering derived by decreasing
Fe-Fe bond length and corresponding MSRD (discussed later).

The EXAFS amplitude depends on several factors and could be given by
the following general equation\cite{Konings}:

\begin{equation}
\chi(k)=
\sum_{i}\frac{N_{i}S_{0}^{2}}{kR_{i}^{2}}f_{i}(k,R_{i})
e^{-\frac{2R_{i}}{\lambda}}
e^{-2k^{2}\sigma^{2}}
sin[2kR_{i}+\delta_{i}(k)]\nonumber
\end{equation}

Here N$_{i}$ is the number of neighboring atoms at a distance R$_{i}$. 
S$_{0}^{2}$ is the passive electrons reduction factor,
f$_{i}$(k,R$_{i}$) is the backscattering amplitude, $\lambda$ is the
photoelectron mean free path, and $\sigma_{i}^{2}$ is the correlated
Debye-Waller (DW) factor, measuring the mean square relative
displacements (MSRD) of the photoabsorber-backscatterer pairs.  Apart
from these, the photoelectron energy origin E$_{0}$ and the phase
shifts $\delta_{i}$ should be known.

We have used conventional procedure to analyze the EXAFS signal
\cite{Konings} due to three shells, i.e., Fe-As, Fe-Fe and Fe-RE
scatterings.  Except the radial distances R$_{i}$ and the corresponding
DW factors $\sigma_{i}^{2}$, all other parameters were kept fixed in
the least squares fit (S$_{0}^{2}$=1).  The EXCURVE9.275 code was used
for the model fit with calculated backscattering amplitudes and phase
shift functions \cite{excurve}.  The number of independent data
points, N$_{ind}\sim$(2$\Delta$k$\Delta$R)/$\pi$ \cite{Konings} was 16
for the present analysis ($\Delta$k=11 \AA$^{-1}$ (k=3-14\AA$^{-1}$)
and $\Delta$R=2.5 \AA).  Starting parameters were taken from the
diffraction studies
\cite{JZhaoNatP,HuangPRB,JZhaoPRB,DCruzNat,Fratini,Margadonna,Karolina,Ozawa}.
 A representative three shell model fit is shown with experimental
Fourier transform as inset to the Figure 2.

The average radial distances as a function of rare earth atom are
shown in Figure 3.  There is a gradual decrease of the average Fe-Fe
and Fe-RE distances (two upper panels) with decreasing rare earth
size, consistent with the diffraction studies showing decreasing
lattice parameters (the a-axis and c-axis as a function of the rare
earth atom are shown as insets) \cite{Ozawa}.  On the other hand, the
Fe-As distance (lower middle panel) does not show any appreciable
change with the rare-earth atom size, revealing strongly covalent
nature of this bond. 
Within experimental uncertainties this appears to be consistent with 
the diffraction results\cite{JZhaoNatP,HuangPRB,JZhaoPRB,
DCruzNat,Fratini,Margadonna,Karolina,CHLee}.

Using the bond lengths measured by EXAFS, we can determine directly
the opening angle at the top of the Fe$_{4}$As tetrahedron (Fe-As-Fe
angle $\theta_{3}$), considered to be the key to the superconductivity
in these materials \cite{JZhaoNatP}.  The Fe-As-Fe angle $\theta_{3}$
has been calculated using the formula; $\theta_{3}$=$\pi$-2cos$^{-1}$
($\frac{d_{Fe-Fe}}{\sqrt{2}d_{Fe-As}}$).  The Fe-As-Fe angle
$\theta_{3}$ is shown in Fig.3.  The Fe-As-Fe angle $\theta_{3}$ is
consistent with the earlier studies, revealing perfect Fe$_{4}$As
tetrahedron \cite{CHLee,JZhaoNatP} for the SmOFeAs.

Figure 4 shows the correlated DW factors as a function of rare earth
atoms measuring the MSRD of different bond lengths.  The MSRD of the
Fe-Fe (middle panel) and Fe-RE pairs (upper panel) appear to depend on
the rare-earth size, while we could hardly see any change in that of
the Fe-As pairs indicating again the stiffness of the later.  The MSRD
of the Fe-Fe shows a clear decrease with decreasing rare earth size,
as the Fe-Fe bond length (Fig.  3).  
Incidentally, the MSRD of the Fe-RE appears to show a small increase 
with decreasing rare earth
size, albeit the change is smaller than that of the Fe-Fe bond length
(upper panel).

Recently Tyson et al \cite{Tyson} have reported temperature dependence
of Fe-As MSRD for the doped and undoped LaOFeAs, showing
that, while the Einstein frequency of the Fe-As mode does not
change with doping, there is a small decrease of static contribution
to the MSRD. The results of Tyson et al \cite{Tyson} are consistent
with strongly covalent nature of the Fe-As bond length.  On the other
hand, the same authors have shown increased Einstein frequency with
doping for the Fe-Fe pair, indicating enhanced Fe-Fe correlations.  

In summary, we have measured the local structure of REOFeAs with
variable rare-earth ion (RE) revealing highly covalent nature of the
Fe-As bond length.  In addition, the Fe-Fe and Fe-RE local atomic
correlations show a systematic change with the rare earth ion size,
evidenced by MSRD of the respective bond lengths.  Considering the
conventional superconductivity mechanism in the strong coupling limit
\cite{McMillan,Santi}, the electron-phonon interaction parameter is
inversely proportional to the phonon frequency, i.e., proportional to
the MSRD (the zero point motion dominates at low temperature and hence
$\sigma^{2}\approx{\hbar}/{2\omega_{E}m_{r}}$, where m$_{r}$ is the
reduced mass and $\omega_{E}$ is the Einstein frequency of the pair). 
Since the T$_c$ of the REOFeAs (if doped) increases with decreasing
rare earth size, it is reasonable to think that the Fe-Fe phonon modes
may not have a direct role in the superconductivity (Fe-Fe MSRD
decrease with decreasing rare-earth size). 
In contrast, the Fe-RE
MSRD tends to show a small increase (or remains constant) with
decreasing rare earth size, and may be somehow contributing to the
superconductivity, however, more experiments are needed to address
this issue.  Although, it is difficult quantify role of local
eletron-phonon coupling in the correlating magnetism and
superconductivity, the presented results certainly provide timely
experimental information on the local atomic fluctuations, that could
be important feed-back for new models to describe fundamental
properties of the REOFeAs with doping and chemical pressure.

\section*{Acknowledgments}
The authors thank the ESRF staff for the help and cooperation during
the experimental run.  We also acknowledge Zhong-Xian Zhao (Beijing)
for providing high quality samples for the present study, and Antonio
Bianconi for stimulating discussions and encouragement.  One of us
(DM) would like to acknowledge $\prime$La Sapienza$\prime$ University
of Rome for the financial assistance and hospitality.  This research
has been supported by COMEPHS (under the FP6 STREP Controlling
mesoscopic phase separation).\\



\begin{figure}
\begin{center}
\includegraphics[width=120mm]{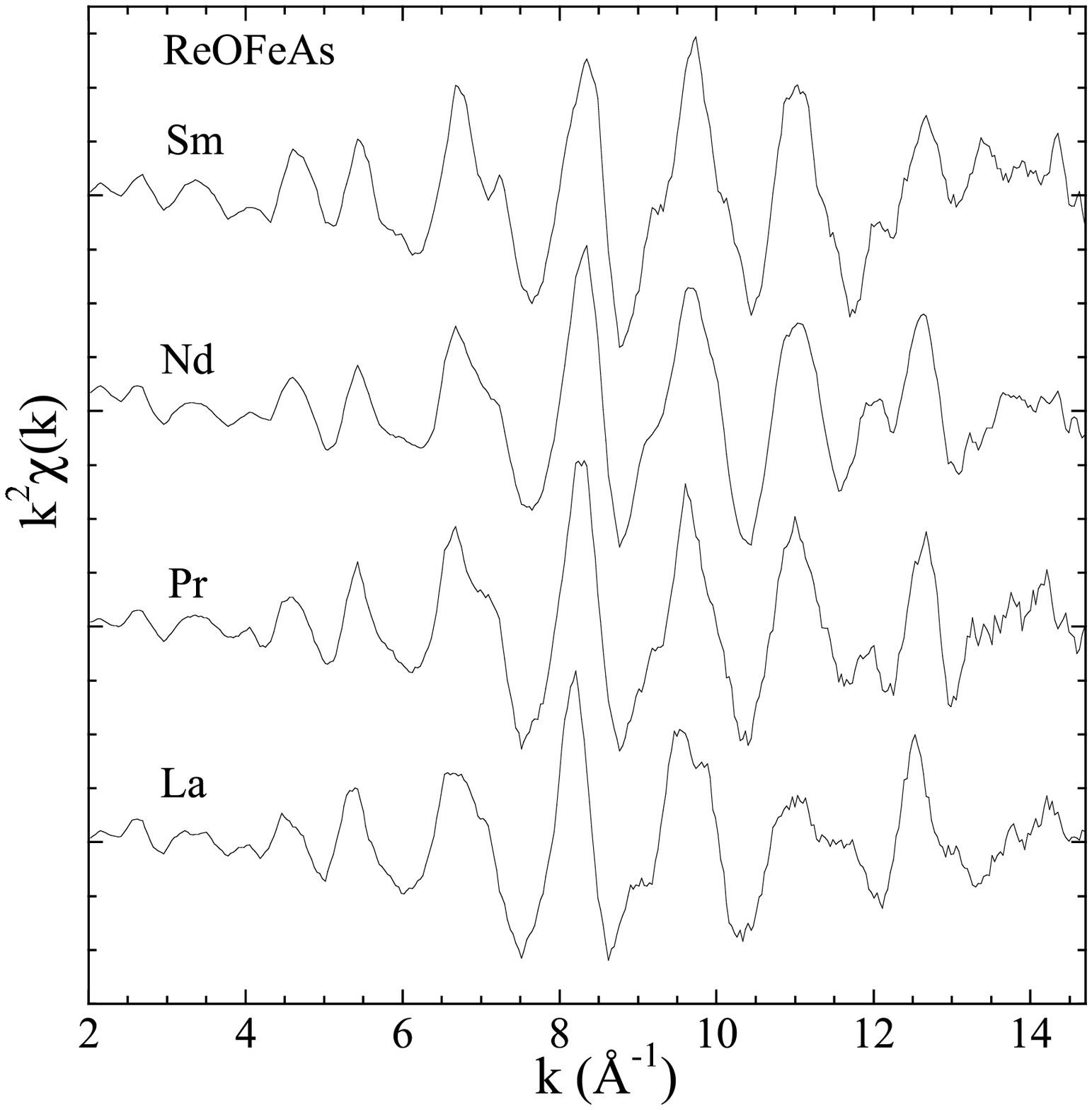}
\caption{\label{fig:epsart} EXAFS oscillations (multiplied by k$^{2}$)
extracted from the Fe K-edge absorption spectra measured on REOFeAs
system (RE =La, Pr, Nd, Sm) at low temperature (40 K) and corrected
for the fluorescence self-absorption effect.}
\end{center}
\end{figure}

\begin{figure}
\begin{center}
\includegraphics[width=120mm]{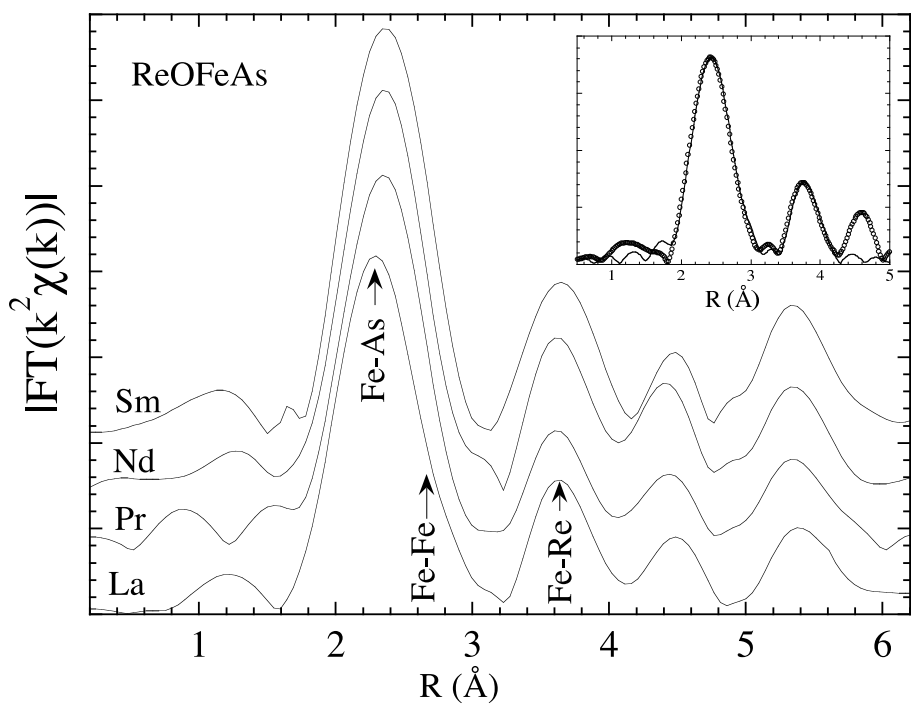}
\caption{\label{fig:epsart}Fourier transforms of the Fe K-edge EXAFS
oscillations showing partial atomic distribution around the Fe in the
REOFeAs system.  The Fourier transforms are performed between
k$_{min}$=3 $\AA^{-1}$ and k$_{max}$=14 $\AA^{-1}$ using a Gaussian
window. The peak
positions do not represent the real distances as the FTs are not
corrected for the phase shifts.  The inset shows a phase corrected
Fourier transform (symbols) (for the LaOFeAs) with a fit over three
shells, i.e., Fe-As, Fe-Fe and Fe-RE.}
\end{center}
\end{figure}

\begin{figure}
\begin{center}
\includegraphics[width=120mm]{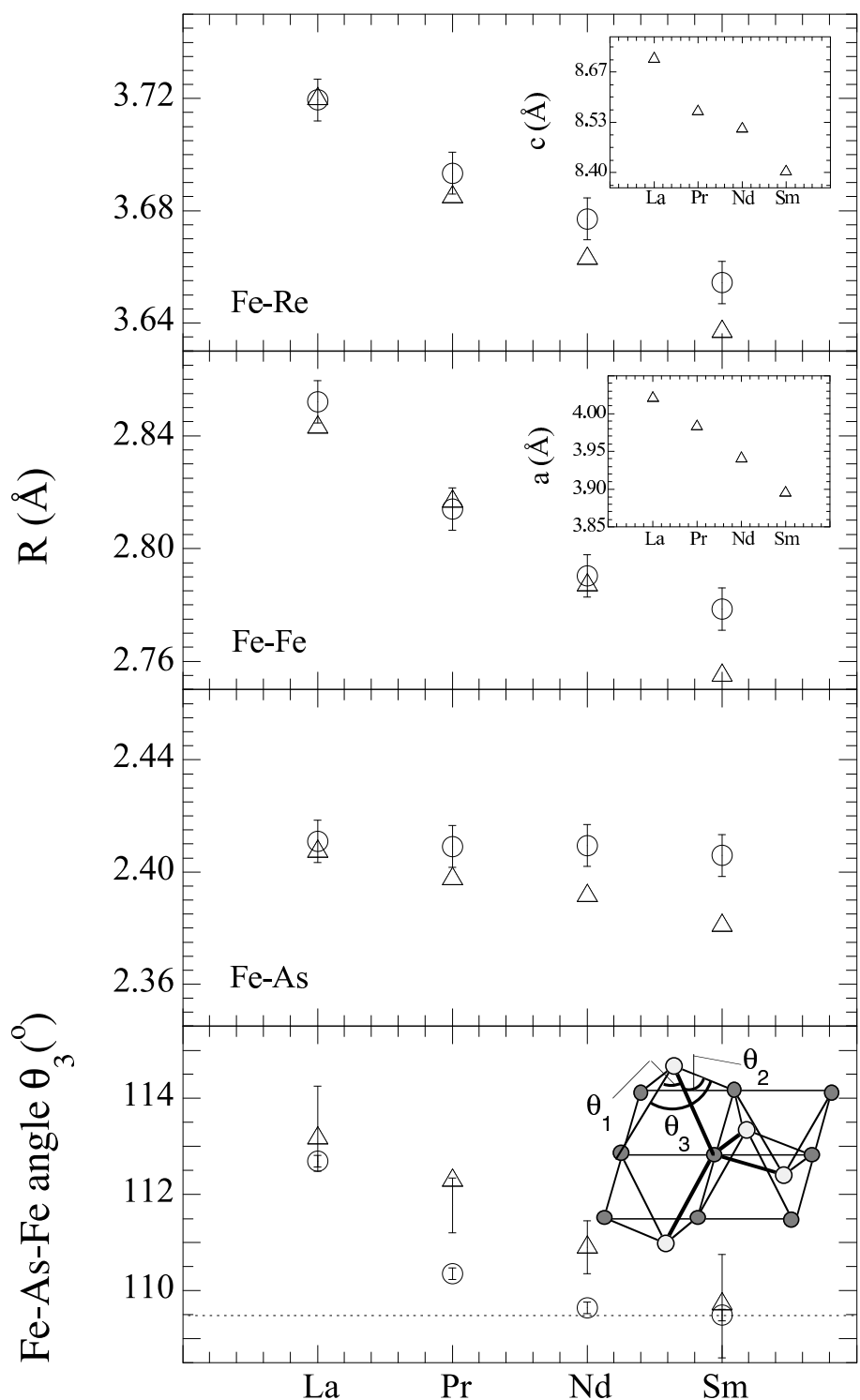}
\caption{\label{fig:epsart} Fe-As (lower middle), Fe-Fe (upper middle)
and Fe-RE (upper) distances at 40 K as a function of rare-earth atom. 
While the Fe-As bond lengths hardly show any change, the Fe-Fe and
Fe-RE bonds change with rare earth size.  The insets (two upper 
panels) show shrinkage of
the lattice parameters with the decreasing rare-earth size 
\cite{Ozawa,Fratini}. The bond 
lengths derived from the diffraction are included in the three 
panels for the comparison. Error
bars represent the average uncertainties estimated by creating
correlation maps.  The Fe-As-Fe angle $\theta_{3}$ determined using
the EXAFS data (circles) is also shown (lower) with the dotted line at
$\theta_{3}$=109.5$^{o}$ corresponding to a perfect tetrahedron. 
The $\theta_{3}$ determined from the diffraction data is shown for 
comparison (triangles). The vertical bars in the lower panel represent 
the span of the 
$\theta_{3}$ measured by diffraction experiments on the 
REOFeAs 
\cite{Ozawa,CHLee,JZhaoNatP,Fratini}. The 
inset shows the cartoon picture of Fe-As-Fe angles\cite{CHLee,JZhaoNatP}.}
\end{center}
\end{figure}

\begin{figure}
\begin{center}
\includegraphics[width=120mm]{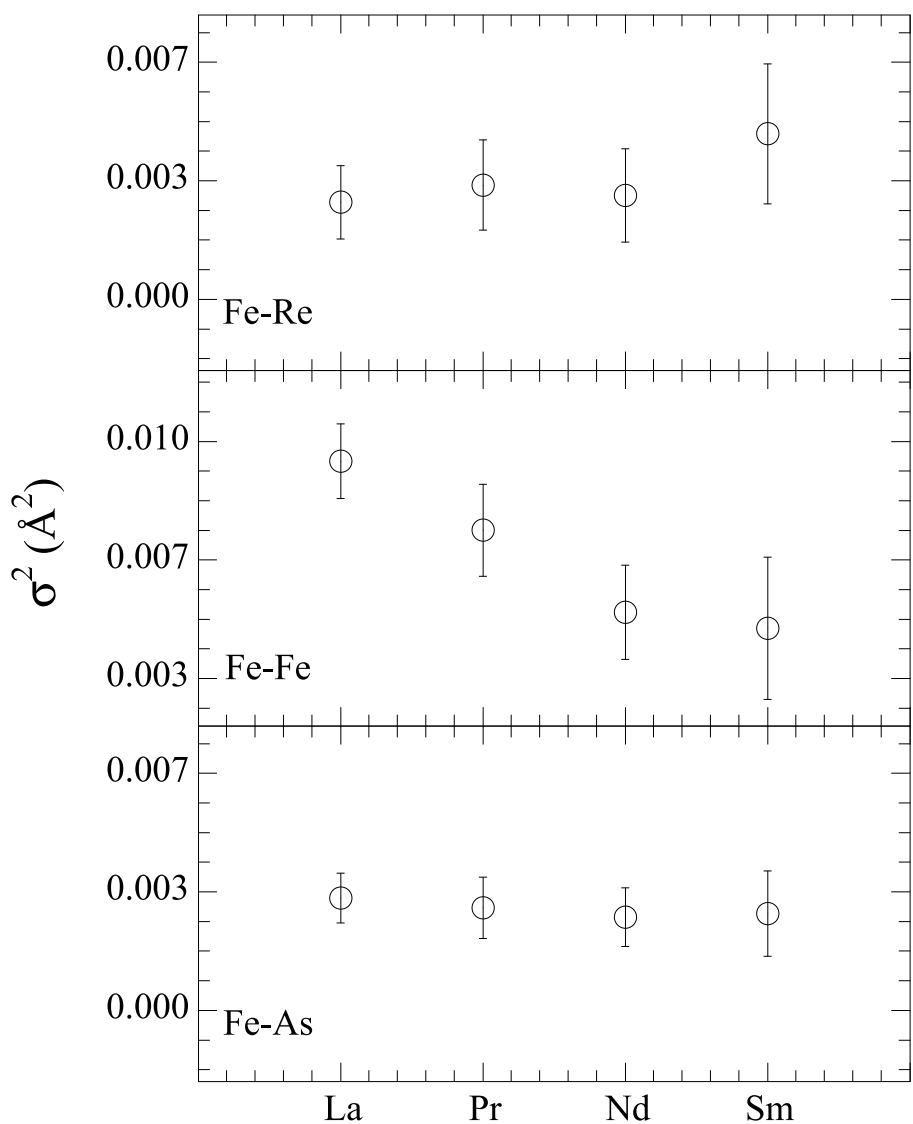}
\caption{\label{fig:epsart}Mean square relative displacements (MSRD)
of the Fe-As (lower), Fe-Fe (middle) and Fe-RE (upper) pairs at 40 K
as a function of rare-earth size.  As the bond length (Fig.  3), the
MSRD of Fe-As hardly show any change indicating strongly covalent
nature of this bond.  
On the other hand, the MSRD of Fe-Fe shows a
decrease while that of the Fe-RE tending to increase from LaOFeAs to
SmOFeAs.}
\end{center}
\end{figure}

\end{document}